\documentclass{evn2002}
\usepackage{graphicx}
\usepackage{astrobib}
\begin{document}
   \title{Towards Proper Motions in the Local Group}

   \author{A. Brunthaler\inst{1}
          \and
	  M. Reid\inst{2}
	  \and
          H. Falcke\inst{1}
	  \and
	  L.J. Greenhill\inst{2}
	  \and
	  C. Henkel\inst{1}
          }

   \institute{Max-Planck-Institut f\"ur Radioastronomie, Auf dem H\"ugel 69, 53121 Bonn, Germany
         \and
            Harvard-Smithsonian Center for Astrophysics, 60 Garden Street, Cambridge, MA 02138, USA 
             }

   \abstract{
Key and still largely missing parameters for measuring the mass content
and distribution of the Local Group are the proper motion vectors of its
member galaxies. The problem when trying to derive the gravitational
potential of the Local Group is that usually only radial velocities are
known, and hence statistical approaches have to be used. The expected
proper motions for galaxies within the Local Group, ranging from 20 to
100 $\mu$as/yr, are detectable with VLBI using the phase-referencing
technique. We present phase-referencing observations of bright masers in
IC~10 and M33 with respect to background quasars. 
We observed the H$_2$O masers in IC10 three times over a period of two
months to check the accuracy of the relative positions. The relative
positions were obtained by modeling the interferometer phase data for
the maser sources referenced to the background quasars. The model
allowed for a relative position shift for the source and a single
vertical atmospheric delay error in the correlator model for each
antenna. The rms of the relative positions for the three observations is
only 0.01 mas, which is approximately the expected position error due to
thermal noise.
Also, we present a method to measure the geometric distance to M33.
This will allow re-calibration of the
extragalactic distance scale based on Cepheids. The method is to measure
the relative proper motions of two H$_2$O maser sources on opposite sides
of M33. The measured angular rotation rate, coupled with other
measurements of the inclination and rotation speed of the galaxy, yields
a direct distance measurement.
   }

   \maketitle
%
%________________________________________________________________

\section{Introduction}

An important astrophysical question is the nature and existence of
dark matter in the Universe, which had been inferred originally from
the flat rotation curves of galaxies (e.g. \citeNP{FichTremaine1991}). The
closest places to look for dark matter halos are the Milky Way and
Andromeda galaxies in the Local Group. Various attempts have been made to 
weigh
the galaxies in the Local Group and determine size and mass of the
Milky Way and its not very prominent dark matter halo 
(\citeNP{KulessaLynden-Bell1992}; \citeNP{Kochanek1996}). Other attempts 
use Local Group
dynamics in combination with MACHO data to constrain the universal
baryonic fraction (\citeNP{SteigmanTkachev1999}).
 
The problem when tyring to derive the gravitational potential of the
Local Group is that usually only radial velocities are known and hence
statistical approaches have to be used. \citeN{KulessaLynden-Bell1992}
introduced a maximum likelihood method which requires only the
line-of-sight velocities (\citeNP{HartwickSargent1978}), but it is
also based on some assumptions (eccentricities, equipartition).
 
Clearly, the most reliable way of deriving masses is using orbits,
which require the knowledge of three-dimensional velocity vectors
obtained from measurements of proper motions. The usefulness of proper
motions was impressively demonstrated for the Galactic Center where
the presence of a dark mass concentration (presumably a black hole, see \citeNP{MeliaFalcke2001}) has been umambiguously
demonstrated by stellar proper motion measurements (\citeNP{EckartGenzel1996};
\citeNP{GhezKleinMorris1998}).
 
However, measuring proper motions of members of the Local Group to
determine its mass is difficult. For the LMC \citeN{JonesKlemolaLin1994} claim
a proper motion of $1.2\pm0.28$ mas/yr obtained from comparing
photographic plates over a timespan of 14 years.  
\citeN{SchweitzerCudworthMajewski1995} claim $0.56\pm0.25$ mas/yr for the 
Sculptor dwarf
spheroidal galaxy from plates spanning 50 years in time. \citeN{Kochanek1996}
shows that inclusion of these marginal proper motions can
already significantly improve the estimate for the mass of the Milky
Way, since it reduces the ambiguity caused by Leo I, which can
be treated as either bound or unbound to the Milky Way. The same work
also concludes that if the claimed optical proper motions are true,
the models also predict a relatively large tangential velocity of the
other statellites of the Galaxy.
The dynamics of nearby galaxies are also important to determine the
solar motion with respect to the Local Group to help define a standard
inertial reference frame. 
 
Despite the promising start, the disadvantage of the available optical
work is obvious: a further improvement and confirmation of these
measurements requires an additional large time span of many decades
and will still be limited to only the closest companions of the Milky
Way.
%__________________________________________________________________

\subsection{Proper Motions with the VLBA}

On the other hand, the expected proper motions for galaxies within the
Local Group, ranging from 1 mas/yr to 20 $\mu$as/yr, are relatively
easy to see with VLBI using the phase-referencing technique. A good
reference point is the motion of Sgr A* across the sky at a speed of 6
mas/yr reflecting the Sun's rotation around the Galactic Center at a
speed of about 220 km/s. This motion is well detected between epochs
separated by only one month with the VLBA (\citeNP{ReidReadheadVermeulen1999}).
 
With the accuracy obtainable with VLBI one could in principle measure
very accurate proper motions for most Local Group members within less
than a decade. The main problem so far is finding appropriate radio
sources. Useful sources would be either compact radio cores or strong
maser lines associated with star forming regions. Fortunately, in a
few galaxies bright masers are already known. Hence the task that lies
ahead of us, if we want to significantly improve the Local Group
proper motion data and mass estimate, is to make phase-referencing 
observations with respect to
background quasars of known Local Group galaxies with strong H$_2$O
masers. 
 
\subsection{Useful Local Group galaxies}
The most suitable candidates for such a VLBI phase-referencing
experiment are the strong H$_2$O masers in IC 10 ($\sim$ 10 Jy peak
flux density in 0.5 km/s line, the brightest known extragalactic maser;
\citeNP{BeckerHenkelWilson1993}) and IC 133 in M33 ($\sim2$ Jy, the first
extragalactic maser discovered). Both masers have been observed
successfully with VLBI~(e.g. \citeNP{ArgonGreenhillMoran1994}, 
\citeNP{GreenhillMoranReid1993}). Additional fainter masers also exist in 
M33 that could be used to extend and improve the studies (e.g. for
constraining galactic rotation; see section~\ref{sec5}).
 
The two galaxies belong to the brightest members of the Local Group
and are thought to be associated with M31. Their line-of-sight
velocities are $-344$~km/s and $-180$~km/s respectively and are
located at a distance of about 800 kpc. In both cases a relatively
bright phase-referencing source is known to exist within a degree.  In
addition their galactic rotation is well known from HI
observations. Consequently, M33 and IC 10 seem to be the best known
targets for attempting to measure Local Group proper motions with the
VLBA.

\section{Observations} 

We observed the H$_2$O masers in IC~10 three times with the VLBA on 2001 
February 09, 2001 March 28 and 2001 April 12 under good weather conditions. 
We observed four 8 MHz bands,
each at right and left circular polarization. The 128 spectral channels
in each band yield a spectral resolution of 62.5 kHz, equivalent to 
0.84 km/s. 

The observations involved rapid switching between two compact extragalactic 
background sources and the H$_2$O masers in IC~10. The source J0027+5958
was taken from the VLBA calibrator survey and was used as the phase-reference
source. Its flux density at 22 GHz was $\approx$ 200 mJy and the angular 
separation on the sky between IC~10 and J0027+5958 is 1$^\circ$.
The second source, J0021+5911, had a flux density of $\approx$ 10 mJy at 22 GHz
and is separated by only 8' from IC~10.
We went through the cycle 
J0027+5958 -- IC10 -- J0027+5958 -- J0021+5911 -- J0027+5958 and the sources 
were changed every 30 seconds. The total observation time was 10 hours each.

The data were calibrated and imaged with standard techniques using 
the AIPS software package. A priori amplitude calibration was applied using 
system temperature measurements and standard gain curves. A fringe fit was 
performed on J0027+5958 and the solutions were applied to IC~10 and J0021+5911.
The phase corrections for all stations except St. Croix showed only slow 
variations with time and it was easy to connect the phases.
The data from the antenna on St. Croix were flagged in all observations due to 
the bad quality of the data. 

\section{Results}

  \begin{figure}
   \centering
   \includegraphics[bb=00 380 600 800,width=10cm,clip]{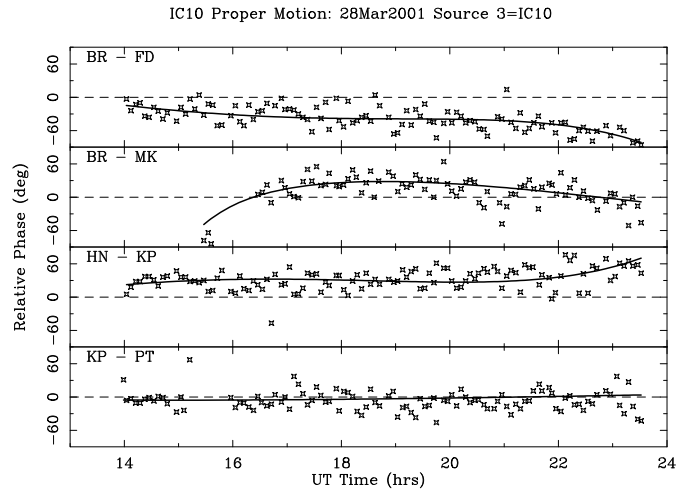}
      \caption{Observed residual phases (stars) and model phase (line) for the 
baselines Brewster -- Fort Davis, Brewster -- Mauna Kea, Hancock -- Kitt Peak 
and Kitt Peak -- Pie Town. The model allows for a relative position offset and a single vertical atmospheric delay.}
         \label{phase_w_a}
 \end{figure}

   \begin{figure}
   \centering
   \includegraphics[bb=00 380 600 800,width=10cm,clip]{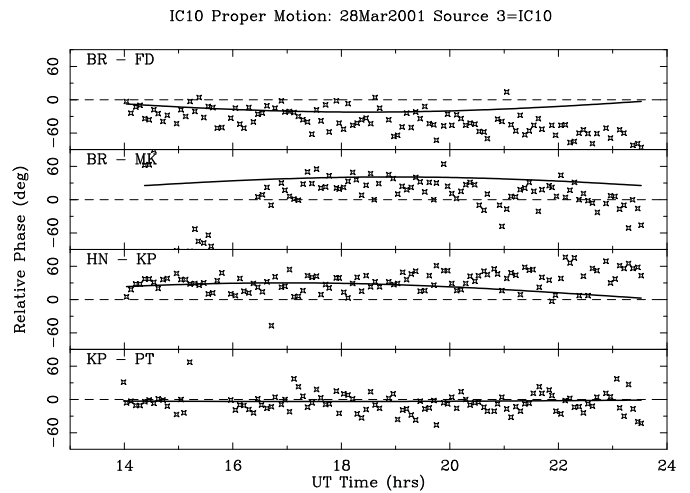}
      \caption{Same as Fig.~\ref{phase_w_a} but the model allows for a relative position offset only.}
         \label{phase_n_a}
 \end{figure}

The three observations within two months were made to check the accuracy and
the repeatability of our results. The second background source was included to
verify the results and to reduce systematic errors. The positions of the 
masers in IC~10 relative to the phase-reference source J0027+5958 were first 
obtained from a model fit to the map for all epochs. The differences in the 
positions between the observations were $\approx$ 0.1~mas. 

The most likely source of relative position error is a small error in the atmospheric model used by the VLBA correlator (see discussion in \citeNP{ReidReadheadVermeulen1999}). To improve our relative position measurements, we modeled our differenced-phase data for the J0027+5958 minus IC~10 pair. The model allowed for a relative position offset for IC~10 and a single vertical atmospheric delay error in the correlator model for each antenna. Fig. \ref{phase_w_a} shows the residual phase of IC~10 and our best model fit for four typical baselines. One can see that the data and the model are in good agreement. Fig.~\ref{phase_n_a} shows the same plot with a model that allows for a relative position shift only. Here the agreement between data and model is much worse.

The results of the position offsets from the model which includes the atmospheric corrections are presented in Table \ref{results}. The deviations of the relative position for the three observations decreased significantly and the rms is now $\approx$ 10~$\mu$as.
The values of the vertical atmospheric delay error were typically of the order of a few cm. These are similar to the values in \citeN{ReidReadheadVermeulen1999}.

   \begin{table}
      \caption[]{Residual positions of IC~10 relative to J0027+5958.}
         \label{results}
     $$
         \begin{array}{p{0.3\linewidth}cc}
            \hline
            \noalign{\smallskip}
            Date   & $East~Offset$ & $North~Offset$ \\
	           & $[mas]$ & $[mas]$ \\
            \noalign{\smallskip}
            \hline
            \noalign{\smallskip}
             2001/02/09 & 0.034& 0.103\\
             2001/03/28 & 0.039& 0.120\\
             2001/04/12 & 0.041& 0.106\\
	     \hline
	     Mean       & 0.038 \pm 0.003& 0.110 \pm 0.007\\
            \noalign{\smallskip}
            \hline
         \end{array}
     $$

   \end{table}

\section{Local Group Proper Motions}
The proper motion vectors of the observed masers will consist of
various contributions, which we list in the following (with numbers for
M33 roughly valid also for IC 10):
 
\begin{itemize}
\item[a)] solar motion around the Galactic Center: 220 km/s ($\sim$~60 $\mu$as/yr at a distance of 800 kpc times the sine
of the angle between the solar motion vector and the line-of-sight to
the galaxy);
\item[b)] motion of M31 with respect to the Milky Way: predicted
to be $\sim$~60 km/s (16 $\mu$as/yr);
\item[c)] orbital motion of M33 (or IC 10) around M31: 100-200~km/s (25-50 $\mu$as/yr);
\item[d)] internal galactic rotation: 110 km/s (30 $\mu$as/yr);
\item[e)] velocity of individual maser components: $\sim$~20-50 km/s ($\sim$5-13
$\mu$as/yr).
\end{itemize}
 
The proper motion vectors (a-c) depend on the mass of M31
and the Milky Way and the distance to the galaxies. These are parameters
we want to constrain with our observations.  The galactic rotation (d)
is very well determined from HI observations. For the intrinsic
velocity of maser components relative to the Local Standard of Rest (e) we 
note that though individual maser components can have velocities of several 
tens of km/s, the average velocity (79 components in IC 133) is just a 
few km/s and hence negligible.
 
Depending on how the vectors add up, we can conservatively expect a
measurable motion of 50-100~$\mu$as/yr. With an astrometric
precision of 10-20~$\mu$as a 5$\sigma$ proper motion result should be
achieved within a year.  Roughly half of this detectable motion is due
to the astrophyscially interesting vectors (a-c). Vector (a), for
example, simply contains a secular parallax which is, however, not
independent of the mass of Andromeda and the Milky Way (vectors b \&
c). For a simple model of the Local Group with two dominant galaxies
and a fixed mass ratio between Milky Way and Andromeda (usually
assumed to be 1:1.5) our two proper motion measurements would already
significantly constrain any mass model for the Local Group. In a more
elaborate approach, which takes the full data set of radial and
proper motions in the Local Group into account (see \citeNP{Kochanek1996}),
the constraints would be even stronger leading to significantly
improved mass estimates and distances.
 
\section{A Direct Measurement of the Distance to M33}
\label{sec5}
Currently, the calibration of most
standard candles used for extragalactic distances,
are tied in one way or another to the distance to the
Large Magellanic Cloud (LMC) (e.g. \citeNP{MouldHuchraFreedman2000}).
However, two relatively new methods
suggest a ``short distance'' to the LMC, about 10\% to 20\%
smaller than from Cepheids.  These methods involve the emission from ``red clump'' stars
using Hipparcos distance calibrations (\citeNP{StanekZaritskyHarris1998}),
and analysis of light and radial velocity curves for the eclipsing
binary HV~2274 (\citeNP{GuinanFitzpatrickDewarf1998}, \citeNP{UdalskiPietrzynskiWozniak1998}).
If the short distance is correct, then indirectly measured distances to
other objects would shrink, increasing estimates of H$_0$ by 10 to 20\%.

We want to obtain a {\it geometric} distance to the nearby
galaxy M33 to permit an independent re-calibration of
extragalactic distance indicators.  This will lead not only to more accurate
distances, and corresponding changes in many physical parameters, but it
will lead to revised estimates of the expansion rate and age of the Universe.
M33 is close enough that both primary and secondary distance
indicators may be readily isolated in ground- and space-based observations
with existing instrumentation.

We intend to measure
the angular rotation rate of M~33 by measuring the relative position
of two H$_2$O maser sources (associated with different regions of
massive star formation), that lie on opposite sides of the galaxy.
Owing to the rotation of the galaxy, the northern source moves
roughly westward while the southern source moves
roughly eastward at a {\it relative} speed of about 220 km/s 
(\citeNP{CorbelliSalucci2000}).
By measuring the angular rotation rate, and comparing it to the
known rotation speed and galaxy inclination, the
distance can be obtained.  Measurements over a 1 year period,
each with an accuracy of $10-20~\mu$as,
would yield an uncertainty in the proper motion, converted to a rotation speed,
of $\approx40-80$~km/s
(for a distance of 800 kpc) or a $3-5~\sigma$ detection. Further observations 
over a longer time range would improve the accuracy of the distance ultimately 
to better than 5 \%. At this point, the distance accuracy would probably be 
limited by knowledge of the inclination of the galaxy.

\section{Conclusion}

The first results of our observations of H$_2$O masers in IC~10 and M33 have 
demonstrated the feasibility of 
high-precision astrometry at the 10 $\mu$as level. With this accuracy we 
expect a 5 $\sigma$ detection of the proper motions of IC~10 and M33 within
one year. The H$_2$O masers in M33 were also observed, but the data reduction
has not been finished yet. The observational techniques are nearly 
the same for IC~10 and M33 and so we expect promising results also for M33.

A possible pitfall for such a project is that individual maser components 
could be short lived and lost with time. However, in M33 the stronger maser 
components are known to exist for now two decades. An additional problem could 
arise if the observed phase referencing sources show motion in a core-jet structure
which is unresolved with the VLBA. This could lead to a small apparent shift 
in the position of the phase referencing source which would be misinterpreted as a 
motion of IC~10 or M33. Using two calibrator sources, we are however able to 
exclude such a bias.

With the second and third set of observations in January 2002 and presumably 
in October 2002 we therefore will be able for the first time to detect 
significant proper motions in the Local Group out to 800 kpc.

\bibliography{aamnemonic,brunthal_refs}
\bibliographystyle{aa}

\end{document}